\documentclass[11pt]{article}
\usepackage{moriond,epsfig}

\newcommand{\nex}{{\sc neXus}~}

\begin{document}
\vspace*{4cm}
\title{SCREENING CORRECTIONS IN SIMULATING HEAVY ION COLLISIONS}

\author{T. PIEROG, H. J. DRESCHER, S. OSTAPCHENKO AND K. WERNER}
\address{SUBATECH, 4 rue Alfred Kastler - La Chantrerie - BP 20722, 44307 NANTES cedex 3, FRANCE}

\maketitle\abstracts{One year ago, we presented a new approach to treat hadronic interactions
for the initial stage of nuclear collisions \cite{Hladik:2001zy,Drescher:2000ha}. It is an effective
theory based on the Gribov-Regge formalism, where the internal structure of
the Pomerons at high energies is governed by perturbative parton evolution,
therefore the name \emph{``Parton-Based Gribov-Regge Theory''.} The main improvement
compared to models used so-far is the appropriate treatment of the energy sharing
between the different elementary interactions in case of multiple scattering. It is 
clear that the above formalism is not yet complete. At high energies (RHIC, LHC), 
the multiple elementary interactions (Pomerons) can not be purely parallel, they 
interact. So we introduce multiple Pomeron vertices into the theory.}

\section{Introduction \label{intro}}
The most sophisticated approach to high energy hadronic interactions is the
so-called Gribov-Regge theory \cite{Gribov:fc}. This is an effective field
theory, which allows multiple interactions to happen ``in parallel'', with
phenomenological objects called \emph{Pomerons} representing elementary interactions
\cite{Baker:cv}. Using the general rules of field theory, one may express cross
sections in terms of a couple of parameters characterizing the Pomeron. Interference
terms are crucial, as they assure the unitarity of the theory. 

A big disadvantage of GRT implementations so far is the fact that cross sections 
and particle production are
not calculated consistently: the fact that energy needs to be shared between
many Pomerons in case of multiple scattering is well taken into account when
considering particle production (in particular in Monte Carlo applications),
but not for cross sections \cite{Abramovsky:bw}. 

Another problem is the fact that at high energies, one also needs a consistent
approach to include both soft and hard processes. The latter ones are usually
treated in the framework of the parton model, which only allows the calculation
of inclusive cross sections.

We recently presented a completely new approach \cite{Werner:ze,Drescher:1999js,Drescher:2000ha} for
hadronic interactions and the initial stage of nuclear collisions, which is
able to solve several of the above-mentioned problems. We provide a rigorous
treatment of the multiple scattering aspect, such that questions of energy conservation
are clearly determined by the rules of field theory, both for cross section
and particle production calculations. In both (!) cases, energy is properly
shared between the different interactions happening in parallel. This is the
most important new aspect of our approach, which we consider a first necessary
step to construct a consistent model for high energy nuclear scattering. 

\begin{figure}[htbp]
{\par\hfill \psfig{figure=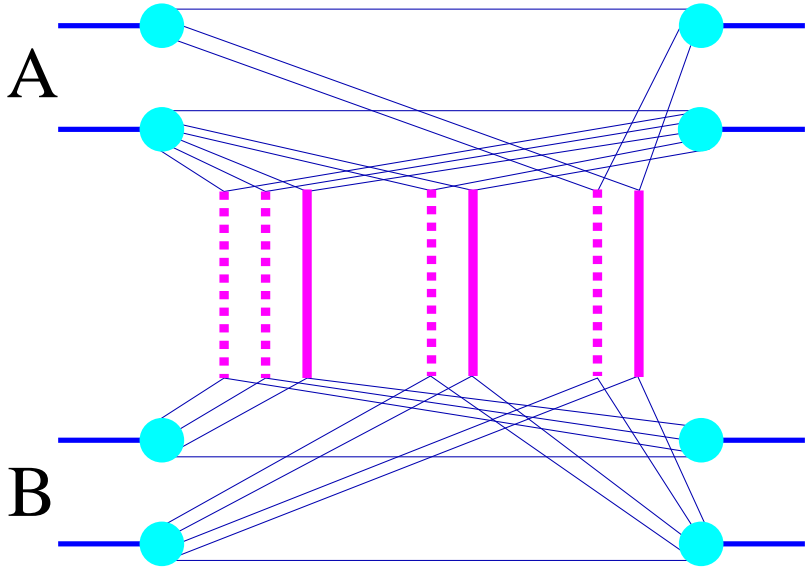,width=11cm} \hfill ~ \par} 
\medskip
\caption{Examples of cut multiple scattering diagrams, with cut (dashed lines) and uncut
(full lines) elementary diagrams (Pomerons). \label{grtppaac}}
\end{figure}

Let us consider nucleus-nucleus (\( AB \)) scattering. The nucleus-nucleus
scattering amplitude is defined by the sum of contributions of diagrams, corresponding
to multiple elementary scattering processes between parton constituents of projectile
and target nucleons. These elementary scatterings are the sum of soft, semi-hard,
and hard contributions: \( T_{2\rightarrow 2}=T_{\mathrm{soft}}+T_{\mathrm{semi}}+T_{\mathrm{hard}} \).
A corresponding relation holds for the inelastic amplitude \( T_{2\rightarrow X} \).
We introduce ``cut elementary diagrams'' as being the sum over squared inelastic
amplitudes, \( \sum _{X}(T_{2\rightarrow X}) \)\( (T_{2\rightarrow X})^{*} \),
which are graphically represented by vertical dashed lines, whereas the elastic
amplitudes are represented by unbroken lines:
\vspace{0.3cm}
{\par  \hfill
\centering \psfig{figure=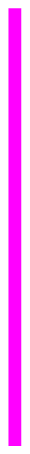,width=0.05cm} \( \quad  \)\( =T_{2\rightarrow 2} \),
\( \quad  \)\( \quad  \)\psfig{figure=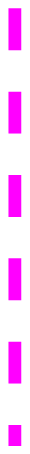,width=0.05cm} \( \quad  \)\( =\sum _{X}(T_{2\rightarrow X}) \)\( (T_{2\rightarrow X})^{*} \). \hfill ~
\par}
\vspace{0.3cm}
\noindent This is very handy for treating the nuclear scattering model. We define
the model via the elastic scattering amplitude \( T_{AB\rightarrow AB} \) which
is assumed to consist of purely parallel elementary interactions between partonic
constituents, described by \( T_{2\rightarrow 2} \). The amplitude is therefore
a sum of many terms. Having defined elastic scattering and inelastic scattering, particle production is practically given, if one employs a quantum mechanically
self-consistent picture. Let us now consider inelastic scattering: one has of
course the same parallel structure, just some of the elementary interactions
may be inelastic, some elastic. The inelastic amplitude being a sum over many
terms -- \( T_{AB\rightarrow X}=\sum _{i}T^{(i)}_{AB\rightarrow X} \) -- has
to be squared and summed over final states in order to get the inelastic cross
section, which provides interference terms \( \sum _{X}(T^{(i)}_{AB\rightarrow X})(T^{(j)}_{AB\rightarrow X})^{*} \).
These can be conveniently expressed in terms of the cut and uncut elementary
diagrams, as shown in fig. \ref{grtppaac}. 

One has to be careful about energy conservation: all the partonic constituents
(lines) leaving a nucleon (blob) have to share the momentum of the nucleon.
So, in the explicit formula one has an integration over momentum fractions of
the partons, taking care of momentum conservation. This formula is the master
formula of the approach, allowing calculations of cross sections as well as
particle production \cite{Drescher:2000ha}.

So far we described only the basic version of the model. In reality we also
consider multiple Pomeron vertices, i.e. enhanced diagrams, which lead to take 
into account screening corrections (next section) and shadowing for nuclear scattering 
(work in progress).

\section{Enhanced Diagrams \label{diagaug}}
\subsection{Presentation}

\begin{figure}[hbtp]
{\par \hfill
\centering \psfig{figure=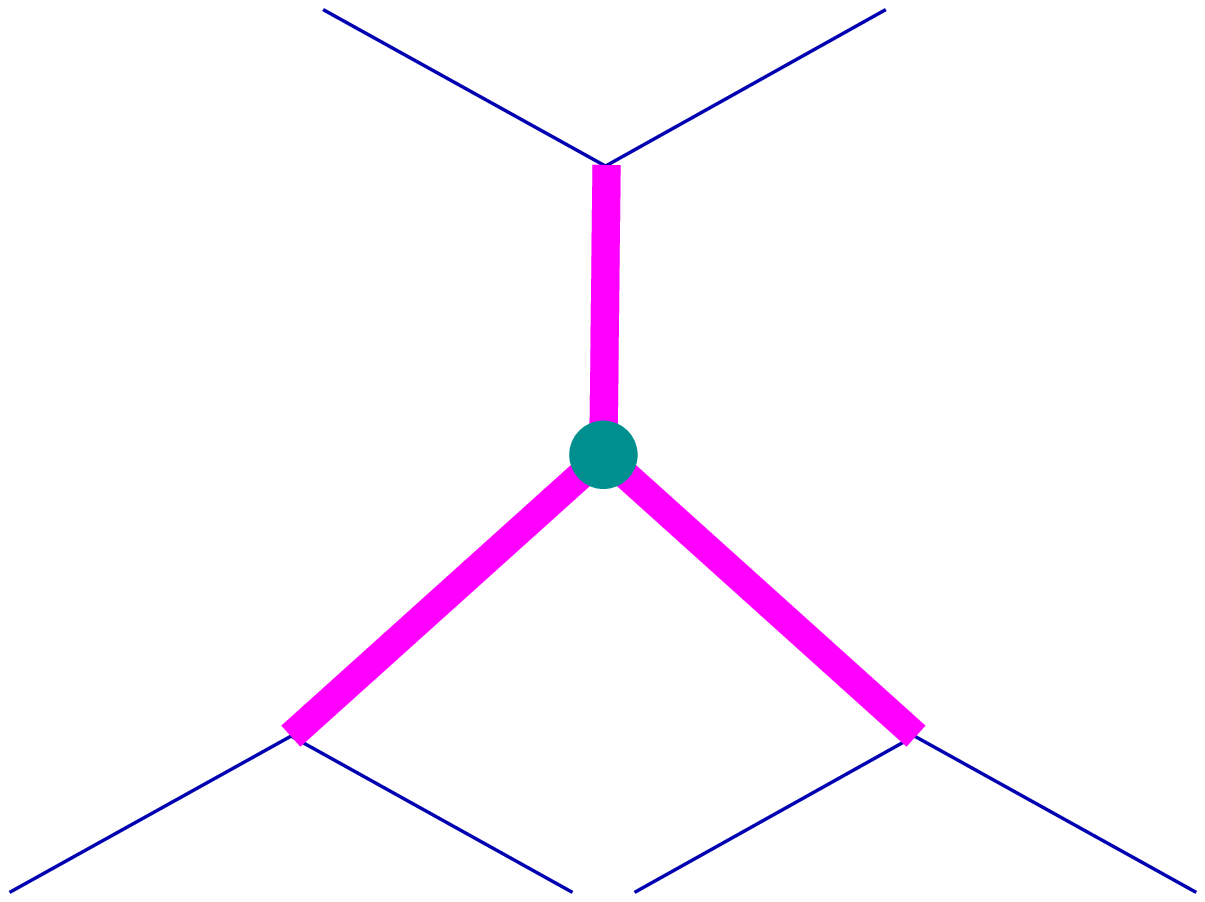,width=2cm,height=2.5cm}\hfill
\psfig{figure=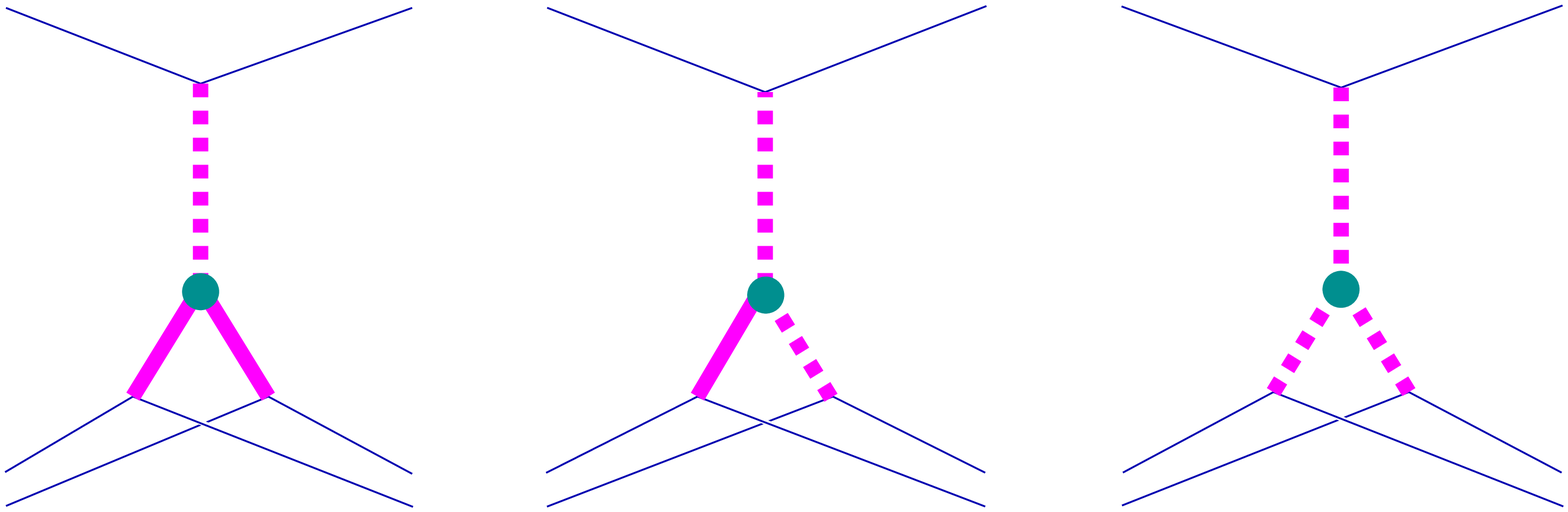,width=5cm,height=2cm}\hfill
\psfig{figure=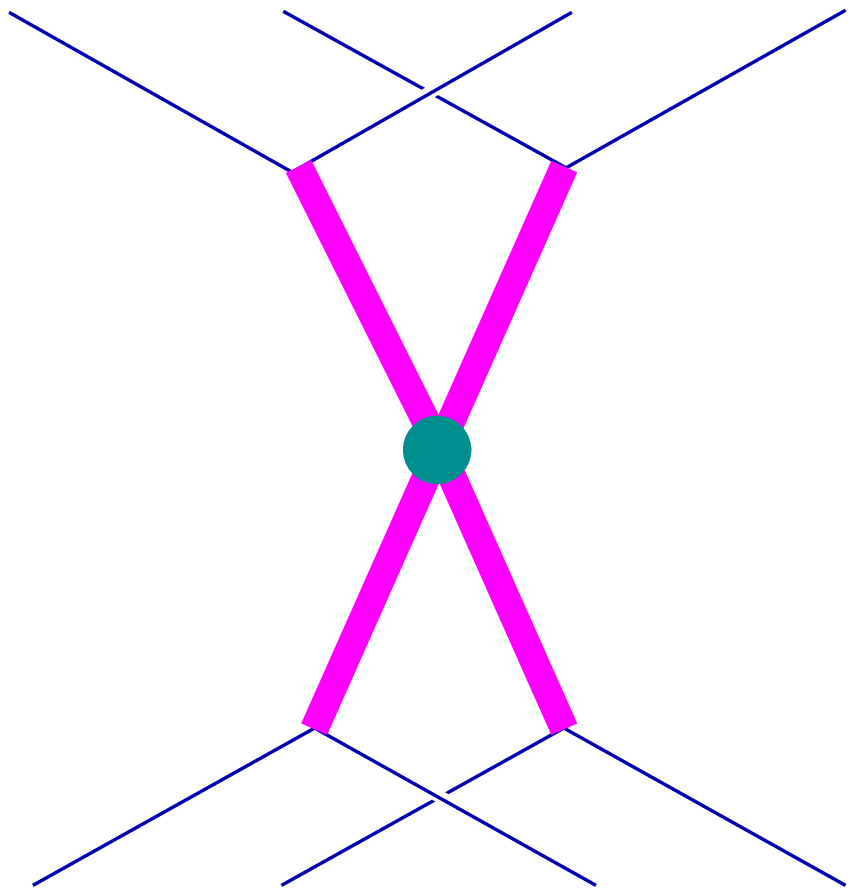,width=2.5cm,height=2.5cm}\hfill
\psfig{figure=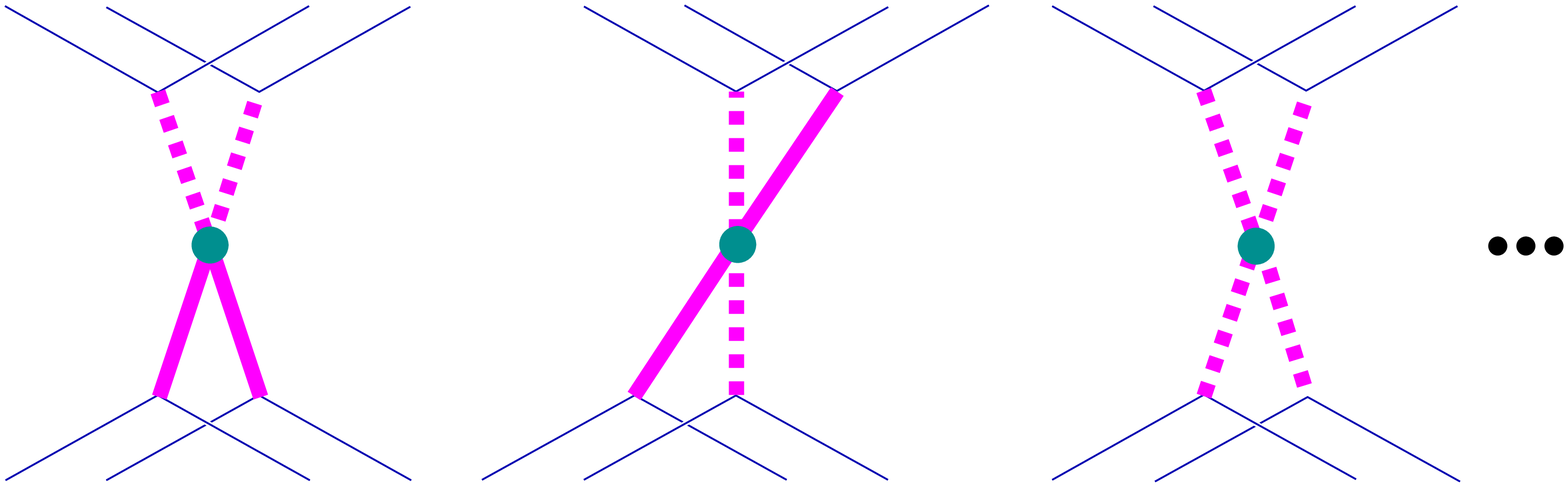,width=5cm,height=2cm}  \hfill ~
\par}
\medskip
\caption{Y uncut and cut diagrams with triple Pomeron vertex and X uncut diagram 
(and some cut ones) with 4-Pomeron vertex.\label{figXY}}
\end{figure}

We saw in section \ref{intro} that a Pomeron is an elementary interaction. 
If those Pomerons interact each other, then they give another type of interaction called 
\emph{enhanced diagram}. There are many types of enhanced diagrams depending on 
the number of Pomeron for each vertices and on the number of vertices. In our model, 
effective first order of triple and 4-Pomeron vertices (Y and X diagrams see fig. 
\ref{figXY}) are enought to be self-consistent.

Enhanced diagrams can be cut by different ways (fig. \ref{figXY}). Each of those 
cut diagrams has a different share to the cross section. For instance, Y diagrams are 
negative contributions, called screening corrections, which lead to a slower increase 
of the total cross section depending on the energy. On the other hand, X diagrams are 
positive contributions to the cross section, i.e. anti-screening. Likewise, diagrams 
do not give the same multiplicity depending on the way they are cut, what leads to 
an increase of the fluctuations.

\subsection{Results for proton-proton scattering}

For the moment, not all enhanced diagrams are included in the \nex model. 
As a consequence, following  
results are calculated only with triple Pomeron diagrams (Y diagrams) with an 
effective coupling constant for pp scattering.

\begin{figure}[hbtp]
{\par \hfill
\centering \psfig{figure=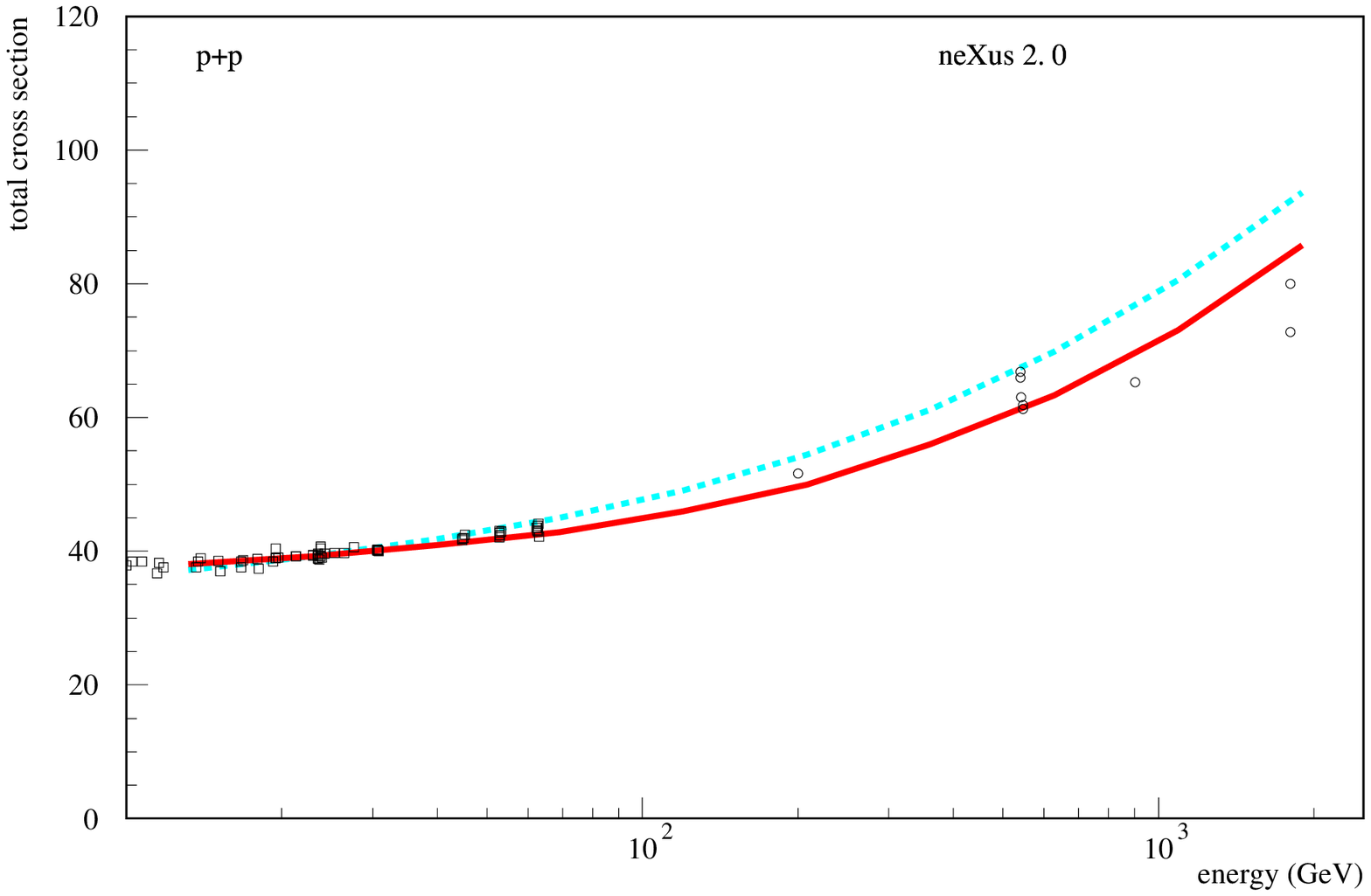,width=7.5cm,height=6.5cm}\hfill 
\psfig{figure=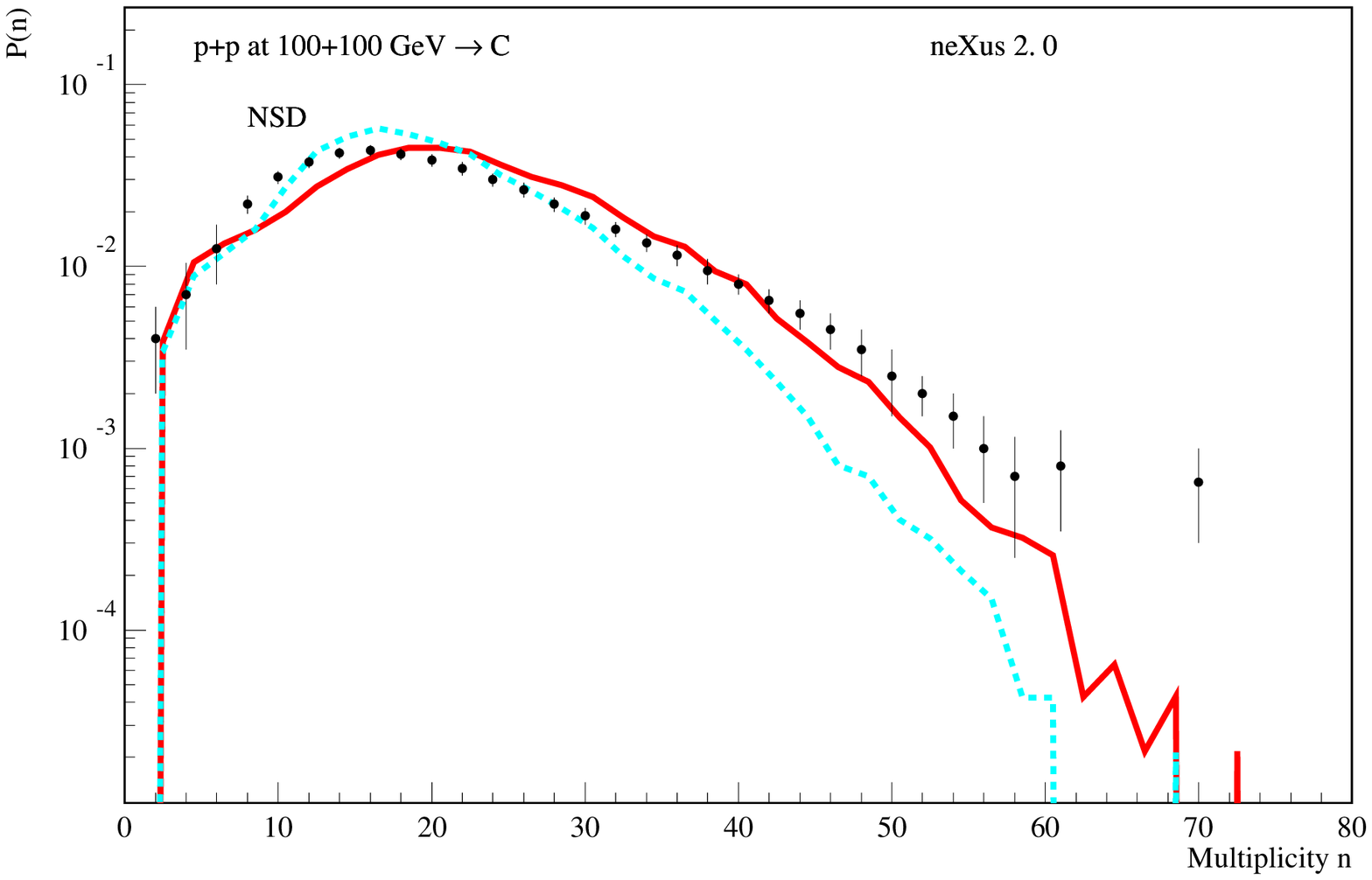,width=7.5cm,height=6.5cm}\hfill ~
\par}
\medskip
\caption{Results for some pp scattering observable : from the left to the right, total cross section versus energy and multiplicity distribution at 200 GeV. The full lines show the results with triple Pomeron and the dash lines whitout. Points are experimental data from \emph{Review of Particle Physics} and \emph{UA5 collaboration}.\label{results}}
\end{figure}

One can observe on the left plot of the fig. \ref{results} a 
deceleration of the growth of the cross section. That means that the interaction's 
probability of proton's components is decreased by screening corrections. 
On the other hand, another consequence of the introduction of enhanced
diagrams is quite visible on the right plot of fig. \ref{results} (probability 
of producing N charged particles). It is noted that the probability of obtaining 
great multiplicities is 5 times larger if one includes triple Pomerons in the model. 
Indeed there are three types of cut triple Pomerons producing a very different 
number of new particles. As 
they are not always the same ones which take place event-by-event, the
fluctuations are increased (sometimes few produced particles,
sometimes much). And thus, on average, it arrives more often than
a lot of particles are produced compared to the basic model.

\section{Conclusion}

The thorough study of the reaction's mechanism of a proton-proton collision
thus showed us that the exchange in parallel of elementary interactions 
called Pomerons with energy conservation 
was not enought to get a self-consistent theory at
high energies. Indeed it is also necessary to consider the interactions
between those Pomerons to have a model which includes the screening corrections. 
One then obtains a more realistic description of the concerned processes. 
Moreover, the introduction of this type of interaction into a nuclear
scattering model allows also to take into account
nuclear effects, such as shadowing effect, which play a major role in
this type of reaction. We thus obtain a realistic and consistent model
which describes the initial stage of ultra-relativistic heavy ions collisions 
and provides the initial conditions necessary to the
description of the following stages of the reaction, 
as an hydrodynamic evolution of a QGP.

\end{document}